\documentclass[openacc]{rstransa}%%%%where rstrans is the template name

\def\titlehead{Perspective}

\begin{document}

% From Science:
% Perspectives discuss one or a cluster of recently published papers or a current research topic of high interest in which an author's perspective sheds an incisive light on key findings in research. These articles typically have one or two authors whose task is to inform our interdisciplinary readership about exciting scientific developments in the author's area of expertise. Other appropriate topics include discussions of methods, books, or meeting highlights. Perspectives are usually between 2000 and 4000 words total (including abstract, main text, references and figure legends). They should have a short pithy title, an abstract of 50 words or less, no more than 35 references, and 1 or 2 figures (with figure legends) or tables.

% Also found this which chimes with what I'm going for:
% Perspectives present a new and unique viewpoint on existing problems, fundamental concepts, or prevalent notions on a specific topic, propose and support a new hypothesis, or discuss the implications of a newly implemented innovation. Perspective pieces may focus on current advances and future directions on a topic, and may include original data as well as personal opinion. These are usually short peer-reviewed articles of around 2000-3000 words. A perspective article usually includes a short abstract of around 150 words and a few tables and figures, if required.

%%%% Article title to be placed here
\title{Machine learning for weather and climate are worlds apart}
% \title{Challenges and opportunities for ML in climate modelling}
% \title{Challenges and opportunities for the emulation of climate models}
% \title{The current state of climate model emulation}
% \title{Data driven models of the Climate}
% 

\author{%%%% Author details
D. Watson-Parris$^{1}$}

%%%%%%%%% Insert author address here
\address{$^{1}$Atmospheric, Oceanic and Planetary Physics, Department of Physics, University of Oxford, UK}

%%%% Subject entries to be placed here %%%%
% \subject{xxxxx, xxxxx, xxxx}

%%%% Keyword entries to be placed here %%%%
% \keywords{xxxx, xxxx, xxxx}

%%%% Insert corresponding author and its email address}
\corres{D. Watson-Parris\\
\email{duncan.watson-parris@physics.ox.ac.uk}}

%%%% Abstract text to be placed here %%%%%%%%%%%%
\begin{abstract}

Modern weather and climate models share a common heritage, and often even components, however they are used in different ways to answer fundamentally different questions. As such, attempts to emulate them using machine learning should reflect this. While the use of machine learning to emulate weather forecast models is a relatively new endeavour there is a rich history of climate model emulation. This is primarily because while weather modelling is an initial condition problem which intimately depends on the current state of the atmosphere, climate modelling is predominantly a boundary condition problem. In order to emulate the response of the climate to different drivers therefore, representation of the full dynamical evolution of the atmosphere is neither necessary, or in many cases, desirable. Climate scientists are typically interested in different questions also. Indeed emulating the steady-state climate response has been possible for many years and provides significant speed increases that allow solving inverse problems for e.g. parameter estimation. Nevertheless, the large datasets, non-linear relationships and limited training data make Climate a domain which is rich in interesting machine learning challenges. 

Here I seek to set out the current state of climate model emulation and demonstrate how, despite some challenges, recent advances in machine learning provide new opportunities for creating useful statistical models of the climate. 

\end{abstract}
%%%%%%%%%%%%%%%%%%%%%%%%%%%

%%%%%%%%%% Insert the texts which can accomdate on firstpage in the tag "fmtext" %%%%%

\begin{fmtext}

% Introduction (brief review)

%% What am I going to talk about
%% Why is it important
%% what have others done in this field
%% How is this different
%% What will I say

% \section{Introduction}

% Fill this up once i have some text

\end{fmtext}

%%%%%%%%%%%%%%% End of first page %%%%%%%%%%%%%%%%%%%%%

\maketitle

\section{Introduction}

Climate models in general, and general circulation models (GCMs) in particular, are the primary tools used for generating projections of climate change under different future socio-economic scenarios. Fully coupled GCMs, which include atmosphere, cryosphere, land and ocean components, are referred to as Earth System Models (ESMs) and are the gold-standard of climate modelling. Due to the large range of spatial and temporal scales and huge number of processes being modelled these are extremely computationally expensive to run and are often only run in coordinated international experiments designed in order to explore particular scientific questions. They also create huge volumes of data which can be difficult to analyse and interpret using traditional tools and methods. There is naturally great interest then in how machine learning (ML) might help to reduce the computational expense in generating this data, or in extracting more value from the data once it is produced~\cite{10.1088/1748-9326/ab4e55,Reichstein_2019}. Here I focus on a third aspect, discussing the current state-of-the-art in climate model emulation for uncertainty quantification and reduction, and highlighting opportunities for new machine learning tools to greatly improve this. 

The need for fast computer simulation emulators has long been recognised in the context of performing inference, where these are often referred to as 'surrogate' models~\cite{ohagan_bayesian_2006}. These surrogates are trained on a few selected samples of the full, expensive simulations using supervised machine learning tools. As a non-linear, non-parametric regression technique, Gaussian processes (GP) are typically used~\cite{doi:10.1111/1467-9868.00294} because of their flexibility and accurate uncertainty estimates. Traditionally, the computational cost of training a GP scales as $\mathcal{O}(N^3)$, where $N$ is the number of training data points, inhibiting their use for large datasets. Recent developments however, have demonstrated new techniques for alleviating these constraints making them competitive with other techniques such as Neural Networks (NNs)~\cite{wang2019exact}. However they are constructed, these surrogate models allow approximating model inversion (determining the inputs given certain outputs) where the exact inverse is not available~\cite{10.2307/2345504}, which is invariably the case for complex models and certainly true for whole climate models. These inverse methods allow the tuning of particular parameters against observations, the analysis and exploration of model uncertainties to different inputs, and the constraint on some of these uncertainties using history matching. 

The uncertainties of GCMs and their output can be broadly categorised in to: 1) Internal variability due to the chaotic fluctuations of the earth system over different time-scales; 2) Model uncertainty due to incomplete or incorrect process representations (structural uncertainty); 3) Model parametric uncertainty due to uncertain input parameters; 4) Scenario uncertainty due to assumptions and incomplete knowledge of the greenhouse gas (GHG) and aerosol and other short-lived climate forcer (SLCF) emissions pathways. 

\begin{figure}[ht]
(a)\centering\includegraphics[width=4.0in]{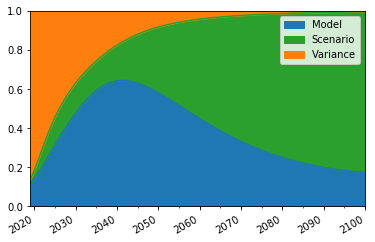}
% (b)\centering\includegraphics[width=2.0in]{CMIP6_od550aer_uncertainty.png}
% \caption{The fractional uncertainty in CMIP6 projections of (a) temperature and (b) aerosol optical depth due to internal variability, model uncertainty and scenario uncertainty as a function of time in to the future.}
\caption{The fractional uncertainty in CMIP6 projections of surface air temperature due to internal variability, model uncertainty and scenario uncertainty as a function of time in to the future.}
\label{fig_CMIP_uncertainty}
\end{figure}

Numerical weather prediction (NWP) models share a common heritage with the atmospheric components of GCMs and are subject to the same uncertainties, however with different emphasis. While in weather prediction the uncertainties in the initial state of the system (1) are a key component, climate projection uncertainties are dominated by model (2+3) and scenario (4) uncertainties over 50 and 100 year timescales respectively~\cite{10.1175/2009bams2607.1,Wilcox_2020}. Figure~\ref{fig_CMIP_uncertainty} shows the fractional uncertainty in the projection of temperature across the CMIP6 multi-model ensemble and demonstrates this clearly\footnote{These uncertainties are calculated as in \cite{10.1175/2009bams2607.1} and described in Appendix~\ref{sec_appendix_a}.}. The internal variability dominates the uncertainty for the first 10 years but rapidly becomes less important as the model, and ultimately scenario uncertainties start to dominate. In exploring climate questions one can thus often neglect internal variability and emulate only the steady state response of the system, significantly simplifying the machine learning problem. 

% Worth pointing this out somewhere:
% When considering these data driven models we primarily discuss the emulating climate models because that's where all of the *projection* data comes from. There are also efforts at using ML for e.g. gap filling historical observations cite(the hadcrut filling paper). 

Quantifying, and ultimately minimising the remaining uncertainties is central to efforts to improve climate projections~\cite{Allen_2002,Collins_2007}, but is also of value when seeking to improve our understanding of the physical climate~\cite{Carslaw_2013}. By framing the discussion of climate emulation around these key uncertainties I hope to demonstrate how machine learning could help in this endeavour. In the rest of this paper I will describe the ways in which climate emulation is already looking to reduce uncertainties in each of the key areas outlined above, before providing an outlook over the ways new and rapidly evolving ML techniques might transform these efforts in the future.

% Current climate emulation
%% Example applications (why do we want emulators)
%% Example tools (how do we currently do it)

\section{Climate emulation}

\subsection{Internal variability}
While short-period (up to a few weeks) internal-variability and uncertainty in the exact current state of the atmosphere dominate uncertainties in weather forecasts, in many climate simulations this is essentially treated as noise which is either controlled for~\cite{Lohmann_Hoose_2009,Jeuken_1996}, or averaged away. In such settings emulating the atmospheric variability is not useful. Longer period, decadal, variability can however be important in climate settings, particularly when comparing historical simulations with observations~[e.g.~\cite{Stevens_2015}]. The use of ensembles of simulations, which sample this uncertainty, enables weather forecast models to generate probabilistic forecasts with improved skill~\cite{Bauer_Thorpe_Brunet_2015} and understand natural variability over climate timescales~\cite{Maher_2019}. These ensembles are extremely computationally expensive to create however, and recent efforts have explored creating machine learning based emulators which could sample this uncertainty more efficiently.

One approach is to emulate the dynamical evolution of these numerical models directly, and this has been explored for both weather~\cite{weatherbench,doi:10.1029/2019MS001705} and climate~\cite{10.5194/acp-20-2303-2020,10.5194/gmd-12-2797-2019,doi:10.1029/2018GL080704}. While these are obviously very early efforts in this direction they demonstrate that developing machine learning models which can compete with their traditional counterparts in numerical weather prediction is extremely challenging, and extending this to climate time-scales even more so, especially given the difficulty in maintaining the energy and mass conservation required for a stable simulation. Where an estimate of the decadal variability is needed, a more promising approach may be to emulate the variability directly from existing ensembles~\cite{Castruccio_2019}. 

\subsection{Model structural uncertainty}

Model uncertainties due to incomplete or incorrect representations of the underlying processes are extremely hard to quantify directly and are often neglected entirely when evaluating individual models against observations. Some estimate can be made by comparing the outputs from multiple models performing the same experiment, often referred to as multi-model ensembles (MMEs), although interpreting any differences is not trivial as many of the models in use around the world are not truly independent and share underlying components~\cite{Knutti_2013}. Further, some models are also demonstrably better or worse in certain aspects~\cite{Pincus_2008} making simple averages over such ensembles potentially misleading. Nevertheless used appropriately, large multi-model ensembles, such as provided by the Coupled Model Inter-comparison Project (CMIP) 5~\cite{Taylor_2011} and CMIP 6~\cite{Eyring_2016} experiments, provide valuable insights in this regard. For example, some early machine learning work in the field developed approaches for combining models from the CMIP5 ensemble~\cite{Monteleoni_Schmidt_Saroha_Asplund_2011}. 

It is worth noting that one of the key ways numerical weather forecasts and regional climate models reduce model uncertainties is by post-processing the predictions using statistical error correction~\cite{Wilcke2013}. A novel approach using ML has recently been proposed for climate models~\cite{Watson_2019} which could provide valuable model improvements, although clearly such an approach can only be validated for observed climate states. 

% # FIrst pass to here
\subsection{Model parametric uncertainty}

The numerical discretization which is necessary to integrate GCMs forward in time defines a spatial (and temporal) scale below which any physical process must be 'parameterized'. These parameterizations are often only approximate representations of the processes they represent and the input parameters must be tuned so as best to reflect the observed climate. There are invariably many combinations of such parameters which can produce a plausible model, a problem termed equifinality~\cite{Beven_Freer_2001}, and so large parametric uncertainty can persist in even the best models. The representation of clouds, for which even the largest examples occur on scales much smaller than typical climate model grid resolutions, is a key uncertainty in this regard~\cite{Stevens_Bony_2013}. Climate feedbacks due to changes in clouds to a given temperature perturbation have been shown to be particularly sensitive to their parameterizations in climate models~\cite{Ceppi_2017}. 

There is a long history of exploring these parametric uncertainties using ensembles of climate simulations sampled across parameter space~\cite{Allen_2000}, including multi-thousand member grand ensembles generated using large networks of home computers~\cite{Stainforth_2005}. Simple linear regression emulators~\cite{Rougier_2007,Sexton:1995bi}, and more recently Gaussian Process (GP)~\cite{ohagan_bayesian_2006} emulators, are then built to span this space so that sensitivity analysis~\cite{Lee:2011796} and parameter inference can be performed by comparison against relevant observations~\cite{Sexton_2019,Watson_Parris_2020}. 

\begin{figure}[ht]
\centering\includegraphics[width=5.0in]{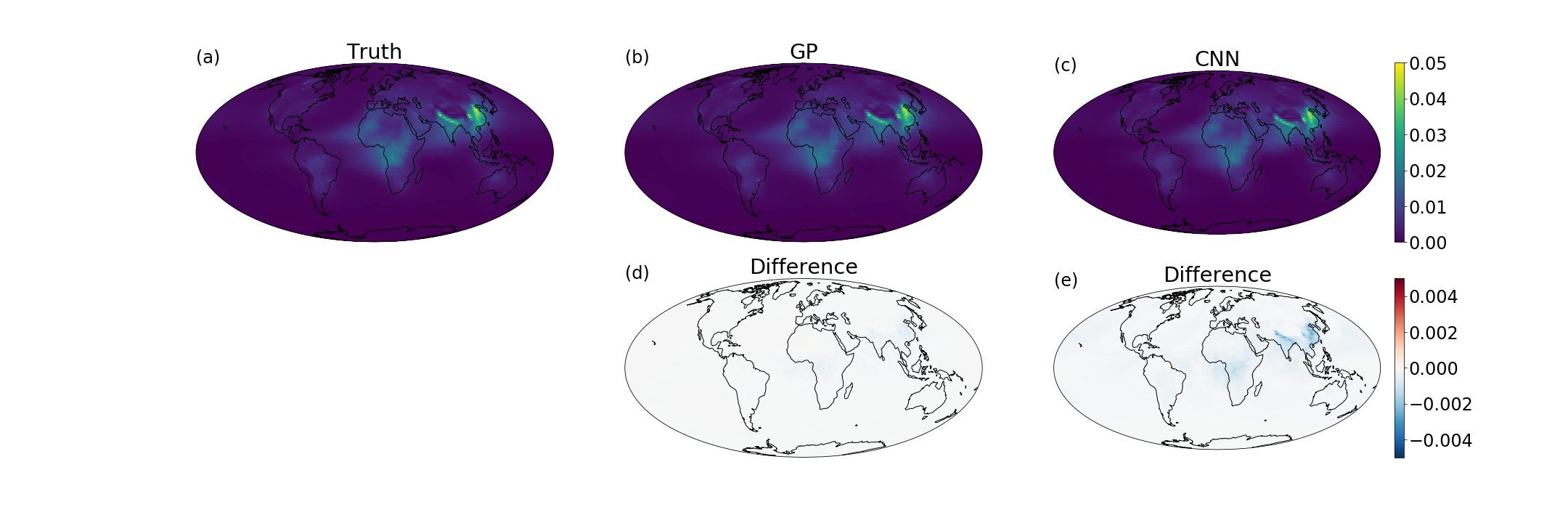}
\caption{The annual mean absorption aerosol optical depth (AAOD) for a particular set of (three) aerosol micro-physical parameters not shown to the emulator during training. a) shows the true modelled output, b) shows the emulated output using a Gaussian Process, c) shows the emulated output using a simple convolution neural network, d-e) show the differences between the modelled data and the Gaussian process emulator and the neural network emulator respectively.}
\label{fig_GP_AAOD}
\end{figure}

An example of an emulator trained on such a perturbed parameter ensemble (PPE) is shown in Figure~\ref{fig_GP_AAOD}. Three parameters identified as being important for the calculation of the absorptivity of aerosol in the atmosphere were perturbed across a wide range of values using a latin hyper-cube sampling. Using a Python package designed to simplify climate model emulation~\footnote{https://github.com/duncanwp/GCEm} the global distribution of Absorption Aerosol Optical Depth (AAOD) is predicted for a particular parameter combination by both a GP and Convolutional Neural Network (CNN) emulator. The errors introduced by emulation are small compared to observation and model-observation comparison errors. This emulator can then be used for comparison against observation to rule out implausible parameter combinations, or infer the optimal set depending on the objective~\cite{gmd-10-1789-2017}. Difficulties in scaling traditional emulators to large datasets and the problem of finding relevant summary statistics has limited their use somewhat and I discuss the opportunities recent advances in ML could provide in the following section.

Machine learning could also be used to completely replace these parameterizations, learning directly from high resolution simulations~\cite{Rasp_2018,Brenowitz_2018} or even observations~\cite{Schneider_Lan_Stuart_Teixeira_2017}. While these can offer some speed improvements they will not drastically decrease the computational expense of running a whole climate model. Indeed, much of their value comes from being able to run \emph{improved} parameterizations, which in turn would lead to better projections (and better training data for whole-model emulators). 

\subsection{Scenario uncertainty}
Over longer time-scales of more than 50 years the scenario uncertainty starts to dominate the model uncertainties. Similarly to the parametric uncertainty discussed above, these uncertainties relate to the inputs of the climate models. The primary distinction is that these input parameters are derived from socio-political considerations, and so cannot be reduced through improved modelling or understanding of the physical climate. Improved sampling of these uncertainties would nevertheless prove valuable to policy makers who need to weigh the cost and impact of different mitigation and adaptation strategies and currently mostly rely on one-dimensional impulse response models~\cite{Meinshausen_Raper_Wigley_2011,Smith_Forster_2018}, or simple pattern scaling approaches~\cite{santer_1990}. Impulse response models are physically interpretable and can capture non-linear behaviour, but are inherently unable to model regional climate changes, while the pattern scaling approaches rely on a simple scaling of spatial distributions of e.g. precipitation by global mean temperature changes, neglecting strong non-linearities in these relationships.

Given the similarity to emulating parametric uncertainty, statistical emulators of the regional climate have been developed~\cite{Holden_2010,Castruccio_2014} although these have been quite bespoke and focus on the relatively simple problem of emulating temperature. Approaches including non-linear pattern scaling~\cite{Beusch_2020} and GP emulation over million-year time-scales~\cite{Holden_2019} hint at the possibility of using modern machine learning tools to produce robust and general emulators over future scenarios. The opportunities, and significant challenges, of realising these possibilities are discussed in the next section.

% I think I can discuss this.

% The output for figure is:\vspace*{-7pt}

% \begin{figure}[!h]
% \centering\includegraphics[width=5.0in]{CMIP6_emulator_paper_v1.1.png}
% \caption{Projected global mean change in surface temperature at 2050 from the present-day emulated from the CMIP6 ScenarioMIP ensemble of models against the cumulative CO2 emitted and change in aerosol optical depth (AOD) due to the different emissions scenarios. Crosses represent training data points and squares are the scenario mean values. The black contour represents 1.2K above present-day (approximately 2K above pre-industrial) temperatures.}
% \label{fig_CMIP6_GP}
% \end{figure}

% \vspace*{-5pt}

\section{Challenges and Opportunities}
Many of the challenges and opportunities which arise in the pursuit of using the plethora of new ML techniques which have recently become available to emulate the climate are common among the potential applications detailed above, and I elucidate some of them below. 

\label{sec_c_and_o}
\subsection{Challenges}
%% (why might it be hard?)

\subsubsection{Few training samples} One of the reasons the latest deep-learning techniques have proved so successful is the enormous number of data samples available for the training of these algorithms. While a single climate model integration can certainly produce many terabytes of data, the numbers of model samples spanning the dimensions over which one might want to emulate is often small. For example, the Community Earth System Model (CESM) Large Ensemble~\cite{10.1175/BAMS-D-13-00255.1} contains only 40 independent members but is over 500Tb in size, as each member represents a long time-series of detailed climate variables. Many multi-model ensembles contain even fewer members. While GPs are well suited to such problems, NNs can easily overfit the limited data. Using neural architecture search~\cite{Kasim_2020} to find the simplest network able to fit the data can help relieve this to some extent. 

\subsubsection{Out of distribution} The training of climate emulators requires an underlying training dataset which spans all possible outcomes to ensure the model does not try and predict outside of the distribution of the training dataset~\cite{npg-26-381-2019}. This requires careful consideration when creating ensembles~\cite{Sexton_2019} and should perhaps be considered when designing future multi-model experiments~\cite{McCollum_2020} to ensure emulators interpolate between training points rather than extrapolate beyond them. Besides well calibrated uncertainties, the use of automatic out-of-distribution detection techniques could prove valuable~\cite{ren2019likelihood}.

\subsubsection{Accurate quantification of uncertainties} Climate model emulation introduces another source of uncertainty in any predictions, and these need to be robustly quantified in order for the prediction to be useful. While GPs provide these by construction, uncertainties of NN predictions can be approximated using dropout~\cite{gal2016dropout}. This is of particular importance given the previous two challenges.

\subsubsection{Short-term and seasonal prediction} Internal variability plays a key role over shorter timescales and cannot be simply averaged away when considering seasonal prediction. Some element of dynamic evolution of the atmospheric state is thus needed in order to accurately emulate these systems, although this can still take the form of simple statistical models of the large scale dynamics~\cite{Cohen_2018}.  

\subsection{Opportunities}

New ML tools and techniques provide opportunities for climate scientists to improve on, and explore new applications for, existing emulators. Besides the important societal impacts of climate research, the large datasets also provides unique opportunities for ML research.

\subsubsection{Large, open datasets} While not always designed to train machine learning emulators, large climate model ensembles from the latest climate models are now available in the cloud, with the tools and infrastructure to easily access them~\cite{Abernathey2017}. This includes ensembles of climate simulations exploring scenario uncertainty~\cite{gmd-9-3461-2016}, model uncertainty~\cite{gmd-9-1937-2016,duncan_watson_parris_2020_3856645}, and natural variability~\cite{10.1175/BAMS-D-13-00255.1}, including some at very high-resolution~\cite{gmd-9-4185-2016}. Training an emulator over combinations of these complimentary ensembles to explore joint uncertainties, or to maximise the available training data, is one promising avenue for further research. This wealth of large spatio-temporal datasets situated next to tremendous computing power also provides opportunities to develop and train more complex emulators.

\subsubsection{New emulators}
To date, emulation has relied on relatively simple techniques on highly aggregated climate data. However, the rapid development of new ML architectures, such as deep GPs~\cite{damianou2013deep,Tran_2016}, Neural Architecture Search~\cite{Kasim_2020} and Spherical-CNNs~\cite{cohen2018spherical} provide exciting opportunities to develop larger, more accurate emulators. These could provide higher spatio-temporal resolution outputs, complementing existing down-scaling techniques~\cite{gmd-9-4087-2016}, or better calibrated uncertainties to account for the large co-variabilities often encountered in climate relevant outputs. 

As described above, due to their huge computational expense climate models are typically run at a coarse resolution, with simplified (parameterized) models used to represent all processes which occur at scales smaller than around $\sim$100km. Previous work has shown how even coarser resolution, computationally cheap, models can be used explore large-scale uncertainty to augment more expensive models ensembles when building emulators~\cite{10.5194/gmd-7-433-2014}. Similarly, such multi-fidelity emulators could be constructed from high-resolution models to encode representations of the unresolved processes, such as clouds and aerosols, using e.g. non-linear auto-regressive GPs~\cite{Perdikaris_2017}. Many of the problems described in this paper also provide challenging settings for ML models and can themselves provide inspiration for new emulators and architectures.

\subsubsection{Improved inference} 
Many current parameter estimation approaches rely on simple rejection sampling to perform model inference, whereby the emulator is sampled from a large number of times and all parameter combinations for which the outputs disagree with observations are rejected. This gradually provides a posterior probability distribution for the input parameters although it requires subjective error metrics and performs poorly for high-dimensional outputs. Simulation based inference is a rich sub-field of machine learning, and many improved techniques are now available~\cite{Cranmer_2020}. Active learning using Bayesian optimization can ensure that training samples are generated where they provide most information for the emulator, and new probabilistic programming tools can use additional diagnostics to improve inference by no longer treating the models as black-boxes. There are also opportunities for automated model calibration and tuning~\cite{cleary2020calibrate} and summary statistic detection to improve the current state-of-the-art.

\subsubsection{Observational emulators} I have primarily focused on the emulation of physical climate models as these are the only tools available for generating future projections. In principle an emulator could be trained on the large satellite based datasets which are now available with the hope that this would provide some skill in future predictions. For example, by training an emulator on observed precipitation and meteorology one could hope to estimate future precipitation changes under a future climate. 

Many significant challenges exist in designing such a system however, in particular the relatively short observational record and the reliance on interpolating in to unknown future states. Encoding strong physical constraints~\cite{beucler2019enforcing} on such a model, for example by enforcing conservation of mass and energy, may provide a useful complement to traditional climate model projections. 

\section{Outlook}
\label{sec_outlook}

%% Ways forward (how do we make the most of the opportunities?)
%%% To make the most of these opportunities we need to combine climate and ML expertise, c.f. knowing the right summary statistics. Cite iMIRACLI and climate informatics, climatechange.ai etc.
%%% Cite K Emmanuel's new paper
%% Conclude

While climate may just be an accumulation of weather, and similar numerical models are used in each domain, as often in the physical sciences more is different\cite{Anderson_1972}. Different processes dominate the responses, different questions are being asked and different uncertainties dominate the predictions. In many respects these differences make climate projections easier to emulate than weather forecasts and much work has been achieved already, but significant opportunities, and some challenges, remain.

The improved techniques available through the recent advances in ML will allow for improved parameter estimation and model tuning; direct emulation of internal variability; emulation of non-linear regional climate responses with higher accuracy and resolution; and potentially observation based models. These will both benefit from, and offer insights into, the underlying physical processes governing our climate.

In order to realise these opportunities we must foster collaborations between the climate and ML communities to develop a shared understanding of the problems and tools available to solve them. Workshops such as this, \url{climatechange.ai} and Climate Informatics (\url{www.climateinformatics.org}) are invaluable in doing so. 

% Even as we endeavour to create models of sufficient complexity and resolution to resolve these issues~\cite{Palmer_Stevens_2019} they will be far too computationally expensive to run large ensembles over internal variability as well as parametric or scenario uncertainty for the foreseeable future.

% Forward modelling the dynamic climate to investigate e.g. transient responses is a much harder ML problem. Energy and mass conservation on decades is crucial for answering questions about what are only small perturbations to the climate system. New tools offer the prospect of achieving this although significant challenges remain.

% The CMIP6 ensemble consists of X experiments and Y model submissions consisting of 20Pb of model diagnostic output. 

% These experiments are often designed to fulfil as many purposes as possible,
% In future, allowing models to 

% [Some of these advances in modelling the steady state could feed back in to emulation of NWP] 

% Climate projections are inherently probabalistic and must include all relevant sources of uncertainty in order to be of most use to policy makers~\cite{Allen_2002}. 

% It is quite possible that experiments at this scale won't be run again and 

\enlargethispage{20pt}

\dataccess{The CMIP6 data used here is available through the Earth System Grid Federation and can be accessed through different international nodes e.g.: https://esgf-index1.ceda.ac.uk/search/cmip6-ceda/. The black carbon PPE data is available here: https://doi.org/10.5281/zenodo.3856644}

\competing{The author declares that they have no competing interests.}

\funding{The author receives funding from the European Union’s Horizon 2020 research and innovation programme iMIRACLI under Marie Skłodowska-Curie grant agreement No 860100 and also gratefully acknowledges funding from the NERC ACRUISE project NE/S005390/1.}

\ack{The author acknowledges the World Climate Research Programme, which, through its Working Group on Coupled Modelling, coordinated and promoted CMIP6. I thank the climate modeling groups for producing and making available their model output, the Earth System Grid Federation (ESGF) for archiving the data and providing access, and the multiple funding agencies who support CMIP6 and ESGF. I also gratefully acknowledge the support of Amazon Web Services through an AWS Machine Learning Research Award.
I thank Mat Chantry for valuable feedback and discussions during the writing of the manuscript.}

\appendix
\section{CMIP6 uncertainty analysis}
\label{sec_appendix_a}

The uncertainty analysis presented in Figure~\ref{fig_CMIP_uncertainty} is calculated using global, annual mean surface air temperature from 20 models that participated in CMIP6 across six scenarios. I follow the approach of \cite{10.1175/2009bams2607.1} but choose not to weight the models since their skill is not of concern, and it makes no significant difference to the results presented here. 

The time-series for each model ($m$) and scenario ($s$) can be represented as:
\begin{equation}
    X_{m,s}(t) = x_{m,s} (t) + i_{m,s} + \epsilon_{m,s}(t)
\end{equation}
where $x$ is a fourth-order polynomial fit using Ordinary Least Squares, $i$ is a reference temperature (taken as the mean between 2015-2020 inclusive) and $\epsilon$ is the residual. The internal variability is assumed to be constant and is defined as the model-mean variance in the residual:
\begin{equation}
    V = |\textrm{var}_{s,t}(\epsilon_{m,s,t})|_m
\end{equation}
The model uncertainty is the scenario-mean variance in the model estimates:
\begin{equation}
    M(t) = |\textrm{var}_{m}(x_{m,s,t})|_s
\end{equation}
while the scenario uncertainty is the variance of the multi-model mean:
\begin{equation}
    S(t) = \textrm{var}_{s}(|x_{m,s,t}|_m)
\end{equation}
The total variance is then the sum of each of these terms:
\begin{equation}
    T(t) = V + S(t) + M(t).
\end{equation}

%%%%%%%%%% Insert bibliography here %%%%%%%%%%%%%%

\bibliographystyle{vancouver}
\bibliography{references.bib}

\end{document}